\documentclass[sigconf]{acmart}

\def\BibTeX{{\rm B\kern-.05em{\sc i\kern-.025em b}\kern-.08emT\kern-.1667em\lower.7ex\hbox{E}\kern-.125emX}}

\usepackage{amsmath} 
\usepackage{todonotes} 
\usepackage{units} 
\usepackage{booktabs} 
\usepackage{fancyvrb} 

\newcommand{\um}[1]{_{\mathrm{#1}}}
\newcommand{\ttt}[1]{\texttt{#1}}
\newcommand{\lmax}{\ell_{\max}}
\newcommand{\numax}{\nu_{\max}}
\newcommand{\nrn}{n_r}
\newcommand{\ket}[1]{\left| #1 \right\rangle}
\newcommand{\bra}[1]{\left\langle#1 \right|}
\newcommand{\braket}[2]{\left\langle\left. #1 \right| #2 \right\rangle}  
\newcommand{\braketR}[2]{\left\langle#1 \left| #2 \right. \right\rangle} 
\newcommand{\braketop}[3]{\left\langle#1 \left. \left| #2 \right| #3 \right. \right\rangle}
\newcommand{\ketbra}[2]{\left|#1 \left\rangle \right\langle #2 \right|}

\begin{document}

\copyrightyear{2019} 
\acmYear{2019} 
\setcopyright{acmlicensed}
\acmConference[PASC '19]{Proceedings of the Platform for Advanced Scientific Computing Conference}{June 12--14, 2019}{Zurich, Switzerland}
\acmBooktitle{Proceedings of the Platform for Advanced Scientific Computing Conference (PASC '19), June 12--14, 2019, Zurich, Switzerland}
\acmPrice{15.00}
\acmDOI{10.1145/3324989.3325717}
\acmISBN{978-1-4503-6770-7/19/06}

%
\title{Analytical PAW Projector Functions \\ for Reduced Bandwidth Requirements}

%

\author{Paul F. Baumeister}
\orcid{0000-0002-2005-4474} 
\affiliation{%
  \institution{J\"ulich Supercomputing Centre \\ Forschungszentrum J\"ulich}
  \streetaddress{}
  \city{J\"ulich}
  \state{Germany}
  \postcode{52425}
}
\email{p.baumeister@fz-juelich.de}

\author{Shigeru Tsukamoto}
\affiliation{%
  \institution{Peter-Gr\"unberg Institut \\ Forschungszentrum J\"ulich}
  \streetaddress{}
  \city{J\"ulich}
  \state{Germany}
  \postcode{52425}
}

%
\renewcommand{\shortauthors}{Baumeister, et al.}

%
\begin{abstract}
Large scale electronic structure calculations require modern
high performance computing (HPC) resources and, as important,
mature HPC applications that can make efficient use of those.
Real-space grid-based applications of Density Functional Theory (DFT)
using the Projector Augmented Wave method (PAW)
can give the same accuracy as
DFT codes relying on a plane wave basis set
but exhibit an improved scalability on distributed memory machines.
The projection operations of the PAW Hamiltonian
are known to be the performance critical part
due to their limitation by the available memory bandwidth.
We investigate on the utility of a 3D factorizable basis of Hermite functions
for the localized PAW projector functions
which allows to reduce the bandwidth requirements for the grid representation
of the projector functions in projection operations.
Additional on-the-fly sampling of the 1D basis functions
eliminates the memory transfer almost entirely.
For an quantitative assessment of the expected memory bandwidth savings
we show performance results of a first implementation on GPUs.
Finally, we suggest a PAW generation scheme adjusted to
the analytically given projector functions.
\end{abstract}

%
\keywords{
Density Functional Theory,
Projector Augmented Wave method,
Real-Space grid,
Cartesian grid,
Radial grid,
Many-core,
GPUs
}

%
\maketitle

\section{Introduction}\label{sec:intro}
%
Uniform real-space grids have proven to be equivalent to plane wave (PW) representations
for electronic structure calculations in the framework of Density Functional Theory
(DFT)~\cite{PhysRev.136.B864,PhysRev.140.A1133}. 
The singularity of the nuclear potential can be represented accurately
on radial grids centered at the atomic site.
Furthermore, rapid oscillations of the quantum mechanical wave functions
caused by the deep attractive potential can be resolved.
The information transfer from 3D Cartesian grids to radial grids and back
is a key ingredient of electronic structure calculations.
Besides other approaches,
an accurate treatment of the singularities
of the electronic potential at the positions of the nuclei
is enabled by the Projector Augmented Wave (PAW) method~\cite{PhysRevB.50.17953}.

The real-space grid approach,
pioneered by Briggs \textit{et al.}~\cite{PhysRevB.54.14362},
is the preferred approach for large scale DFT problems on supercomputers
as it leads to very sparse matrix representations of the Kohn-Sham Hamiltonian
and is free of global Fourier transforms.
These properties allow an improved scalability on parallel computers
compared to calculations involving a PW basis.
Most implementations of real-space DFT
sample the projector functions on a real-space grid
and store these in memory, as implemented for frozen-core PAW~\cite{PhysRevB.71.035109,PhysRevB.82.205115} 
or using pseudopotentials~\cite{MARQUES200360,IWATA20102339}, 
a limit case of PAW.\@ 

Then, the calculation of inner products between localized projector functions
and fully extended Kohn-Sham wave functions requires projection operations
which are typically bounded by the memory bandwidth (BW) of the computing device.
Comparing to the peak achievable rates of floating-point operations,
memory BW has become a scarce resource
on modern high-performance computing systems.

\noindent
In this work, we suggest analytical shapes for the PAW projectors
which can lower the memory BW requirements
for projection and expansion operations
in implementations of the PAW and pseudopotential methods.

\noindent
In this paper we make the following contributions:
\begin{itemize}
 \item We introduce the Spherical Harmonic Oscillator (SHO) as a natural link between 3D Cartesian grids and radial grids.
 \item We assess the potential to accelerate electronic structure calculations using SHO eigenstates.
 \item We analyse the quality of existing PAW projectors when represented in a SHO basis.
 \item We suggest a new PAW generation method adjusted to the analytical projector shapes.
\end{itemize}

\noindent
The rest of this paper is structured as follows.
In sec.~\ref{sec:theory} we introduce general properties of the SHO basis and basics of DFT calculations.
In sec.~\ref{sec:BW} we outline the application of a SHO basis
for PAW projector functions under aspects of high-performance computing.
In sec.~\ref{sec:QLT} the quality of a SHO basis expansion for existing PAW data sets is assessed.
We suggest a modification to the PAW generation scheme in sec.~\ref{sec:MOD}.
In sec.~\ref{sec:PERF} expectable performance improvements on graphics processors (GPUs) are assessed.
Finally, sections~\ref{sec:discuss},~\ref{sec:related} and~\ref{sec:summary}
contain a discussion of this approach, related works and a summary, respectively.

\noindent
Throughout the paper we use atomic Rydberg units: $m_e = \nicefrac{1}{2}$, $\hbar = 1$ and $e^2 = 2$,
i.e.~length is measured in Bohr ($\equiv \unit[.529]{\mbox{\normalfont\AA}}$)
and the unit of energy is Rydberg ($\equiv \unit[13.6]{eV}$).

\section{Theory}\label{sec:theory}

\subsection{Spherical Harmonic Oscillator}\label{sec:SHO}
\noindent
The quantum-mechanical Spherical Harmonic Oscillator (SHO)
\begin{equation}
 \hat H\um{SHO} = -\vec \nabla^2 + \vec r^2 \sigma^{-4}
 \label{eqn:SHO_Hamiltonian}
\end{equation}
is a well-understood problem, often referred to
as three-dimensional isotropic harmonic oscillator.
Here, $\sigma$ is a length scale parameter in Bohr units which also defines the energy scale.
The 3D partial differential equation (PDE) can be solved using separation of variables,
either in the set of Cartesian variables $(x,y,z) \equiv \vec r$
or, due to the spherical symmetry of the potential term,
using spherical coordinates $r$, $\vartheta$ and $\varphi$.
\subsection{Cartesian Coordinates}
For the Cartesian approach it is sufficient to find the solutions
of the 1D harmonic oscillator Hamiltonian
\begin{equation}
 \hat H\um{1D} = -\frac{\partial^2}{\partial x^2} + x^2 \sigma^{-4}
 \label{eqn:1HO_Hamiltonian}
\end{equation}
which leads to the eigenenergy $E\um{1D} = (2n$+$1)\sigma^{-2}$
with the quantum number $n \in \mathbb N_0$.
The corresponding eigenstates $\ket{\psi_{n}}$ are given by
Hermite polynomials times a Gaussian envelope function:
\begin{equation}
 \psi_{n}(x) \propto H_{n}\left( \frac{x}{\sigma} \right) \ \exp(-\frac{x^2}{2\sigma^2})
  \label{eqn:1HO-eigenstate}
\end{equation}
where $H_n(x)$ are Hermite polynomials.
\begin{figure}
	\centering
	\includegraphics[width=\linewidth]{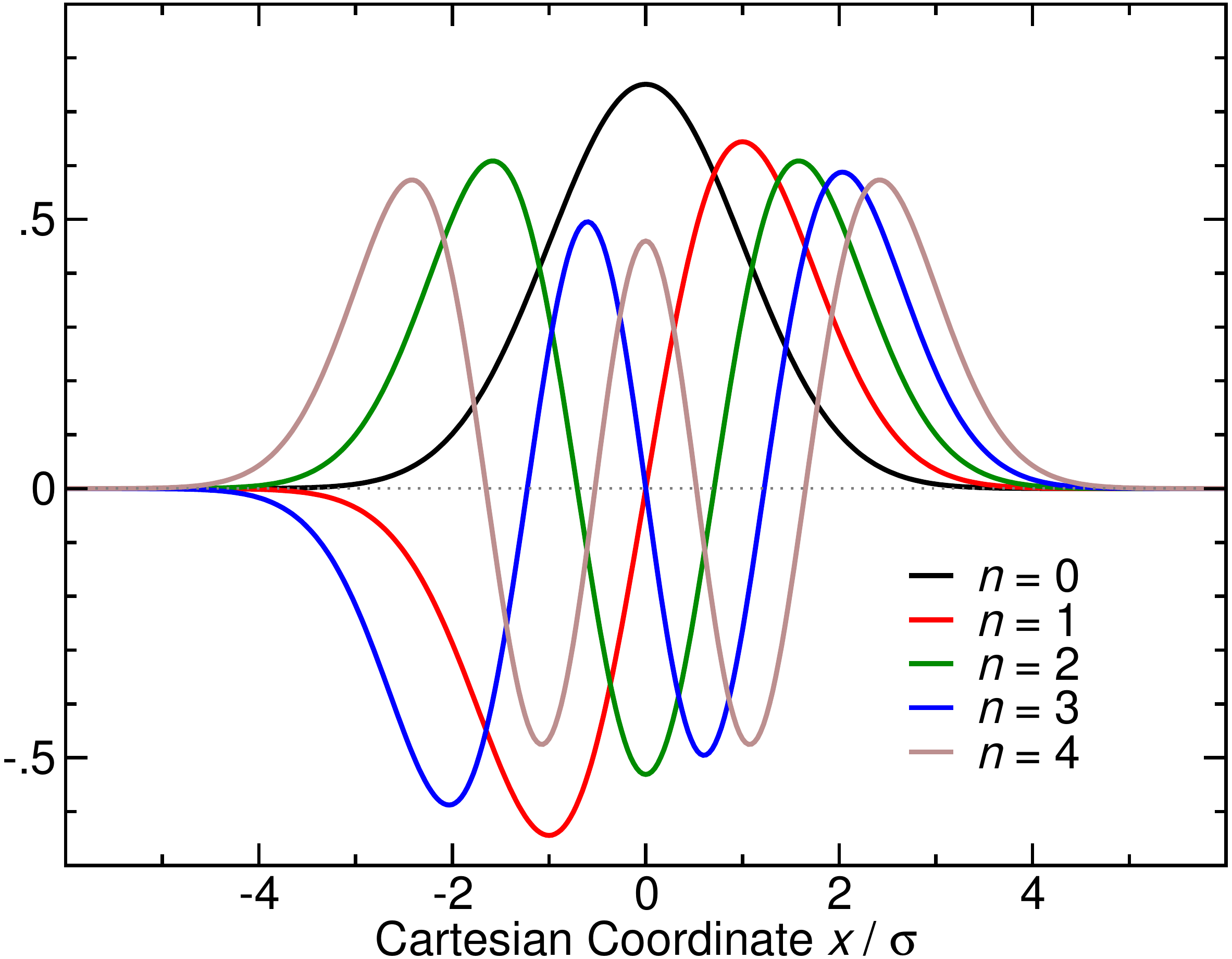} 
    \caption{\label{fig:normalized_1HO_states}
    The eigenstates of the 1D harmonic oscillator are
    Hermite polynomials times a Gaussian, see eq.~(\ref{eqn:1HO-eigenstate}).
   	}%
   	\Description{
   	Set of eigenfunctions of the 1D harmonic oscillator.
   	}%
\end{figure}
Proper normalizing factors are omitted here for simplicity.
The lowest five normalized Hermite functions are shown in fig.~\ref{fig:normalized_1HO_states}.
\\
\noindent
For the 3D problem,
three independent quantum numbers $(n_x,n_y,n_z)$ $\in \mathbb N_0^3$
and the eigenenergy
\begin{equation}
 E\um{SHO}\,\sigma^2 = 2(n_x + n_y + n_z) + 3 \coloneqq 2\nu + 3
 \label{eqn:SHO-energy-Cartesian}
\end{equation}
are found. The eigenstates factorize in the three spatial dimensions:
\begin{align}
 \Psi_{n_x n_y n_z}(x,y,z) &= \psi_{n_x}(x) \, \psi_{n_y}(y) \, \psi_{n_z}(z)
  \label{eqn:SHO-factorizable-basis} \\
  &= H_{n_x} \left( \frac x \sigma \right) \  H_{n_y} \left( \frac y \sigma \right) \
  H_{n_z} \left( \frac z \sigma \right) \ \exp(-\frac{\vec r^2}{2\sigma^2})
  \label{eqn:SHO-eigenstate-Cartesian}
\end{align}

\subsection{Spherical Coordinates}
The PDE in spherical coordinates $(r,\vartheta,\varphi)$ leads to the radial quantum number
$\nrn \in \mathbb N_0$, the angular momentum character $\ell \in \mathbb N_0$
and the magnetic quantum number $m \in [-\ell, \ell] \cap \mathbb Z$:
\begin{equation}
 \Psi_{\nrn \ell m}(r,\vartheta,\varphi) = R_{\nrn \ell} \left( \frac{r}{\sigma} \right)
   \ Y_{\ell m}(\vartheta,\varphi)
   \label{eqn:SHO-eigenstate-radial}
\end{equation}
Here, $Y_{\ell m}$ denotes spherical harmonic functions.
The eigenstates of the differential equation for the radial coordinate $r$ are
\begin{equation}
	R_{\nrn \ell}(r) \propto r^\ell L^{(\ell + \frac 12)}_{\nrn}(r^2) \ \exp(-\frac 12 r^2)
\end{equation}
where $L^{(\alpha)}_n$ are associated 
Laguerre polynomials.
\begin{figure}
	\centering
	\includegraphics[width=\linewidth]{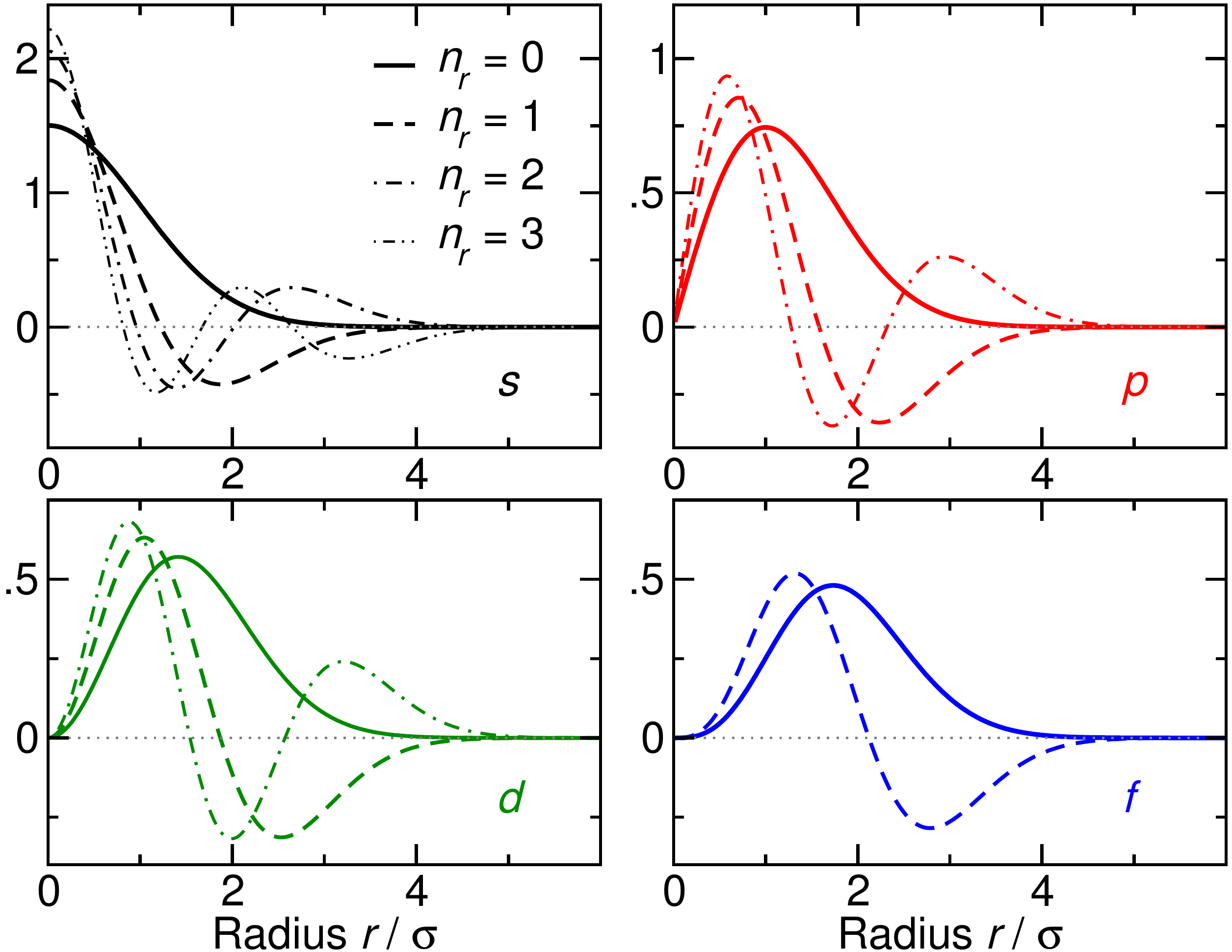} 
    \caption{\label{fig:radial_SHO_states_by_ell}
    Radial part of the lowest eigenstates
    of the spherical harmonic oscillator up to $\numax = 6$ for $\ell \in [0,3]$.
   	}%
   	\Description{
   	Radial eigenfunctions of the spherical harmonic oscillator.
   	All functions behave like $r^\ell$ at the origin.
   	The radial functions are a product of a Gaussian times
   	a polynomial expression in $r$ where the lowest function is nodeless,
   	the next higher function features one node, and so on.
   	}%
\end{figure}
See fig.~\ref{fig:radial_SHO_states_by_ell} for the set of normalized radial functions $R_{\nrn \ell}(r)$.
Due to the spherical symmetry of $\hat H\um{SHO}$ the eigenenergy does not depend on $m$:
\begin{equation}
 E\um{SHO}\,\sigma^2 = 2(\ell + 2\nrn) + 3 \coloneqq 2\nu + 3
 \label{eqn:SHO-energy-radial}
\end{equation}
\begin{figure}
	\centering
	\includegraphics[width=\linewidth]{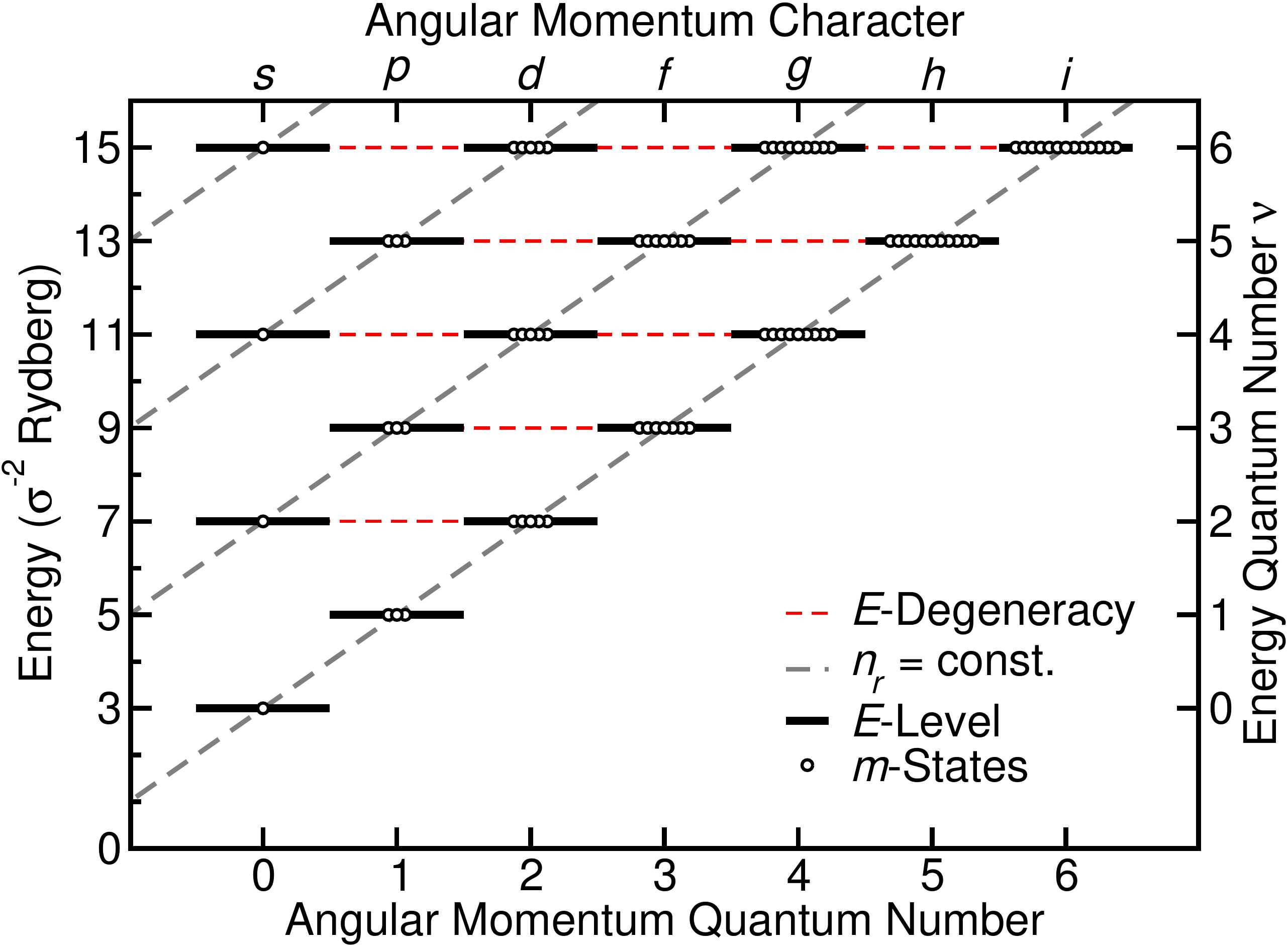}
    \caption{\label{fig:radial_SHO_energies}
    Energy levels of the spherical harmonic oscillator.
    The dashed gray diagonals correspond to a constant number of radial nodes, $\nrn$.
    The lowest diagonal ($\nrn$ = $0$) corresponds to Gaussian type orbitals.
    The red dashed horizontal lines indicate energy-degenerate subspaces indexed by $\nu$.
    }
    \Description{
    Energy level diagram of the spherical harmonic oscillator.
    Levels are spaced by \unit[2]{Rydberg} starting from \unit[3]{Rydberg}.
    From $\ell$=2 on, linear combinations in the energy-degenerate subspace
    lead to the $(x,y,z)$-factorizable representation in Cartesian space.
    }
\end{figure}
An energy diagram for $\nrn$, $\ell$ and $\nu$ is shown in fig.~\ref{fig:radial_SHO_energies}.

\subsection{Connection of Subspaces}
In eq.~(\ref{eqn:SHO-energy-radial}) and eq.~(\ref{eqn:SHO-energy-Cartesian}),
the energy quantum number $\nu \in \mathbb N_0$ was introduced.
As both approaches solve the same problem
the $\frac 12(\nu + 2)(\nu + 1)$ energy-degenerate solutions
in each subspace indexed by $\nu$
must span the same functions space.
In fact, it is possible to find a unitary 
transform $\hat U$
such that
\begin{equation}
 \Psi_{n_x n_y n_z}(x,y,z) \equiv \sum_{\nrn \ell m} U_{n_x n_y n_z}^{\nrn \ell m} \  \Psi_{\nrn \ell m}(r,\vartheta,\varphi)
 \label{eqn:SHO-transform-Cartesian-to-radial}
 \text{.}
\end{equation}
Matrix elements of $U_{n_x n_y n_z}^{\nrn \ell m}$ are only non-zero if
\begin{equation}
	\ell + 2\nrn = \nu = n_x + n_y + n_z \text{.}
\end{equation}
Furthermore, even within each $\nu$-block many entries of $\hat U$ vanish due to other symmetries.

\subsection{SHO Basis Set}
The full set of SHO eigenstates forms a basis of
smooth functions defined on $\mathbb R^3 \mapsto \mathbb C$.
Therefore, for both representations, radial and Cartesian, the completeness relations
\begin{align}
  1 &= \sum^{\infty}_{\nrn \ell m} \ket{\nrn \ell m} \bra{\nrn \ell m} \qquad \text{and} \\
  1 &= \sum^{\infty}_{n_x n_y n_z} \ket{n_x n_y n_z} \bra{n_x n_y n_z}
\end{align}
must hold.
Later, we will exploit that
\begin{align}
  1 &= \sum^{\infty}_{\nrn \ell m} \ket{\nrn \ell m} \bra{\nrn \ell m} \sum^{\infty}_{n_x n_y n_z} \ket{n_x n_y n_z} \bra{n_x n_y n_z} \\
    &= \sum^{\infty}_{\nrn \ell m} \ket{\nrn \ell m} \sum^{\infty}_{n_x n_y n_z} \braket{\nrn \ell m}{n_x n_y n_z} \bra{n_x n_y n_z} \\
    &= \sum^{\infty}_{\nrn \ell m} \sum^{\infty}_{n_x n_y n_z} \ket{\nrn \ell m} U_{n_x n_y n_z}^{\nrn \ell m} \bra{n_x n_y n_z} \coloneqq \hat{\mathcal B}_{\infty}
    \label{eqn:unitary_unity} 
    \text{.}
\end{align}
Truncating the series at a given maximal energy $E_{\max} = (2\numax$+$3)\sigma^{-2}$
will break the completeness relation, $\hat{\mathcal B}_{\numax} \neq 1$.
However, through the energy cut-off criterion it is ensured that
\begin{equation}
   \sum_{ \nrn \ell m }^{ 2\nrn + \ell    \leq \numax } \ket{\nrn \ell m} \bra{\nrn \ell m}
 = \sum_{ n_x n_y n_z }^{ n_x + n_y + n_z \leq \numax } \ket{n_x n_y n_z} \bra{n_x n_y n_z}
   \label{eqn:SHO-basis}
\end{equation}
are both projection operators describing the same subspace of functions.
The size of a finite SHO basis is
\begin{equation}
 N\um{SHO} = \frac 16(\numax+3)(\numax+2)(\numax+1)
 \label{eqn:SHO_basis_size} \text{.}
\end{equation}

\subsection{Electronic Structure Methods}
In the effective potential landscape arising from DFT
two major regions can be identified: core and interstitial.
Close to an atomic \emph{core}
the attractive Coulomb potential dominates.
Here, spherical coordinates $(r,\vartheta,\varphi)$
are best to solve the Kohn-Sham equation.
Radial grids densely sample the radial coordinate $r$, typically in a logarithmic fashion.
All angular dependencies are expanded in spherical harmonics $Y_{\ell m}(\vartheta,\varphi)$.
In the \emph{interstitial} region the potential is strongly determined
by the geometry of neighboring atoms and their ionic character.
Here, we have to solve a 3D PDE in order to find valence states
that describe chemical bonding.

\noindent
There is a large variety of DFT methods that differ
either by the basis set used for the 3D problem
or by the way these basis functions are connected to the radial grids of each atom.

\noindent
DFT methods based on a 3D plane wave (PW) basis set
can make use of 
\begin{equation}
 \exp(\imath \vec k \cdot \vec r) = 4\pi \sum_{\ell = 0}^\infty \imath^\ell \sum_{m=-\ell}^{\ell}
  Y^*_{\ell m}(\hat{\vec k}) \, Y_{\ell m}(\hat{\vec r}) \, j_\ell(|\vec k|\cdot|\vec r|)
  \label{eqn:Rayleigh_equation}
\end{equation}
which yields a natural transition from Cartesian space $\vec r \in \mathbb R^3$
to radial coordinates $r=|\vec r|$
via spherical Bessel functions $j_\ell(x)$
and spherical harmonics $Y_{\ell m}(\vartheta,\varphi)$.
Methods exploiting this equation directly are referred to as Augmented Plane Wave (APW) methods
and implementations of e.g.~the (full-potential) linearized APW methods are
often considered the most accurate electronic structure codes~\cite{Singh:2006,doi:10.1080/10408436.2013.772503}. 
However, a basis consisting of $N\um{PW}$ plane waves
leads to a dense matrix representation of the Kohn-Sham Hamiltonian
requiring $\mathcal O(N\um{PW}^2)$ memory and $\mathcal O(N\um{PW}^3)$ operations
to diagonalize them. 
As this easily exceeds the system capacities of a single compute node,
in particular memory capacity,
we have to consider the efficiency of parallel eigensolvers for dense problems
which typically require intensive all-to-all communication.
\\
A vast number of DFT calculations are performed on PW methods \
using Bl\"{o}chl's PAW method~\cite{PhysRevB.50.17953},
e.g.~\cite{QE-2009,Torrent2008337,PhysRevB.59.1758}. 
PW-PAW can also make use of eq.~(\ref{eqn:Rayleigh_equation})
for the calculation of the overlap between the PWs and
PAW projector functions localized in real-space.
As this also leads to a large dense Hamiltonian matrix,
the sparse representation of the PAW Hamiltonian in real-space is preferred in the limit of large problems.

Switching between a real-space grid and a PW representation of the wave functions
demands for the application of global 3D Fourier Transforms (FFTs)
which are strongly memory-bound operations and
limit the scalability on distributed memory machines.

\noindent
Real-space grid DFT is, by construction, free of FFTs
and can be combined with the PAW method equally well.
Therefore, it is the approach most suited for supercomputing.

\section{Saving Bandwidth}\label{sec:BW}
\subsection{High-Performance Computing}\label{sec:HPC}
The expected availability of Exascale supercomputers
gives hope for the feasibility of large-scale calculations for real materials
using DFT and related methods.
However, conventional representations and solver algorithms
exhibit all-to-all communication patterns
which do not scale well on distributed memory architectures.
Real-space grid based representations have shown to
scale efficiently on large allocations of HPC machines.
Using domain decomposition,
mainly nearest-neighbor inter-process communication
in a periodic 3D topology is required.
This allows to parallelize work with large numbers of tasks efficiently.
However, the evolution of HPC compute nodes
undergoes a strong change from scalar CPUs towards
many-core architectures featuring wide vector units.
In particular, modern GPUs offer
very high floating-point operation performances compared to their power consumption
and can, therefore, be found in many of the worlds largest HPC machines\footnote{top500.org}.
The increase of parallelism due to
more cores,
wider vectors,
increased instruction level parallelism and
extended capabilities for simultaneous multithreading
lets the processing power in terms of arithmetic operations
grow faster over years than technology permits for the memory BW.
Despite the advent of new memory technologies, BW has become
one of the most costly resources
for many applications in the field of computational science
that cannot leverage usual cache optimizations.
This leads to an urgent need for new bandwidth-saving algorithms.

\subsection{Hamiltonian Action}\label{sec:HMT}
For the real-space grid-based PAW approach,
the performance of iterative eigensolver methods, 
like e.g.~ChASE~\cite{Winkelmann:2018:ChASE},
relies on a computationally efficient implementation of the action of
the Hamiltonian operator onto wave functions.
The PAW Hamiltonian takes the form
\begin{equation}
 \hat H\um{PAW} = \hat T + \tilde V\um{eff} + \sum_{aij} \ket{\tilde p^a_{i}} D^{a}_{ij} \bra{\tilde p^a_{j}}
 \label{eqn:PAW_Hamiltonian}
\end{equation}
with the kinetic energy operator $\hat T$
and a smooth local effective potential $\tilde V\um{eff}(\vec r)$.
The largest fraction of compute time for the action of $\hat H\um{PAW}$
onto (preliminary) eigenvectors goes to the sum over the dyadic expressions
$\ketbra{\tilde p_i}{\tilde p_j}$
for each atom $a$.
When applying $\hat H\um{PAW}$
to a given set of independent wave functions, $\ket{\Psi_k}$,
the application of the so-called \emph{non-local} potential part
is performed in three steps:
\begin{enumerate}
 \item Projection: $c^a_{jk} = \braket{ \tilde p_j^a }{ \Psi_k }$
 \item Multiplication: $d^a_{ik} = \sum_{j} D^a_{ij} \, c^a_{jk} $
 \item Expansion:  add $\sum_{ai} d^a_{ik} \ket{ \tilde p_i^a }$ to $(\hat T + \tilde V\um{eff})\ket{\Psi_k}$
\end{enumerate}
In real-space PAW implementations,
the projector functions need to be represented on the real-space grid.
Following the original PAW scheme, these functions are given as
\begin{equation}
 \tilde p^a_i(\vec r) = P^a_{n \ell}(|\vec r_a|) Y_{\ell m}(\hat{\vec r}_a)
\end{equation}
where $\vec r_a = \vec r - \vec R_a$ for each atom $a$ and the index $i$ encodes $(n,\ell,m)$.
In contrast to the spherical harmonics $Y_{\ell m}$
the radial parts $P^a_{n \ell}(r)$ are given numerically
and usually depend only on the atom species of $a$.
In the spirit of linearized APW methods, typical projector set sizes are limited to two projectors per $\ell$
for $s$-, $p$-, and $d$-projectors~\cite{JOLLET20141246}.
Hence, a typical setup used for calculations
involving transition metal atoms comprises six different radial functions $P^a_{n \ell}(r)$
which translates to $18$ projector functions $\tilde p^a_i(\vec r)$ per atom.
The latter number defines the index range for $i$ and $j$ in eq.~(\ref{eqn:PAW_Hamiltonian}).
Light elements typically feature less projectors.
For rare earth elements, adding $f$-projectors is mandatory.

\subsection{Projector Filtering}\label{sec:Filtering}
Following the original PAW methodology~\cite{PhysRevB.50.17953},
projector functions are supposed to be strictly confined to the inside of a
sphere of radius $R\um{aug}$ around the atomic position $\vec R_a$.
For clarity of the notation we suppress the atom index $a$ in the rest of sec.~\ref{sec:BW}.
The atomic spheres are constructed to be non-overlapping
so that $R\um{aug}$ is usually chosen as half the nearest-neighbor distance (touching spheres).
This hard localization constraint leads to
an unphysical aliasing effect (\emph{egg box effect}).
A spurious dependency of the classical DFT observables (total energy and atomic forces)
on the relative position of an atom w.r.t.~the positions of grid points can be measured.
Removing high frequency components from $\tilde p_i(\vec r)$ mitigates this effect
and various methods have been proposed to filter the projector functions in reciprocal space.
Soler and Anglada~\cite{SOLER20091134,PhysRevB.73.115122} summarize the problem
of finding the best localization in both, real-space and reciprocal space.

Many implementations of real-space grid-based DFT make use of the filtering recipe
by Tafipolsky and Schmid~\cite{doi:10.1063/1.2193514}
where a Gaussian-shaped mask function restores the localization in real-space
after suppressing high frequency components in the Bessel-transformed representation
of the radial part of a projector function.

Only the \emph{double-grid technique} by Ono \emph{et al.}~\cite{PhysRevLett.82.5016}
applies a filter to the product $P_{n \ell}(|\vec r|) Y_{\ell m}(\hat{\vec r})$ in 3D space
which implies that also high frequency components that stem from the spherical harmonics
$Y_{\ell m}$ are successfully suppressed.

Nevertheless, one effect of filtering is an increased projection radius, $R\um{prj} > R\um{aug}$,
around the atomic position where the representation of the projector functions on the grid is non-zero.
This strongly adds to the cost of the Hamiltonian action proportional to ${(R\um{prj} / R\um{aug})}^3$.

\noindent
To summarize:
for reasons of costs it is essential to preserve the localization of PAW projector functions in real-space.
For reasons of accuracy also localization in reciprocal space is mandatory.

\subsection{Data Compression}\label{sec:compress}
One goal is to make the Hamiltonian action perform as efficient as possible
on current HPC compute systems, i.e.~vectorized many-core architectures.
The projection operation~(1) and its counterpart, the expansion~(3),
can be written as matrix-matrix multiplications.
Here, one operand is a sparse matrix since the projector functions
are non-zero only close to the atom. 
The sparsity leads to a strong limitation of the performance by the memory BW.

A major strategy for the reduction of BW requirements
is to decrease the communication volume, i.e.~by data compression.
A projector functions represented on a uniform
Cartesian real-space grid with grid points at $\vec \jmath \in \mathbb N^3$ 
can be viewed as rank-3 tensor $P_{j_x j_y j_z}$.
Here, tensor compression methods
as studied intensively by Khoromskaia and Khoromskij~\cite{C5CP01215E} 
try to approximate high-rank tensors by sums of dyadic products.
In this particular case we seek a representation
\begin{equation}
  P_{j_x j_y j_z} \approx \sum_{n_x n_y n_z} \bar P_{n_x n_y n_z} \, v^{[x]}_{n_x j_x} \, v^{[y]}_{n_y j_y} \, v^{[z]}_{n_z j_z}
  \label{eqn:TuckerTensor}
\end{equation}
\begin{figure}
	\centering
	\includegraphics[width=\linewidth]{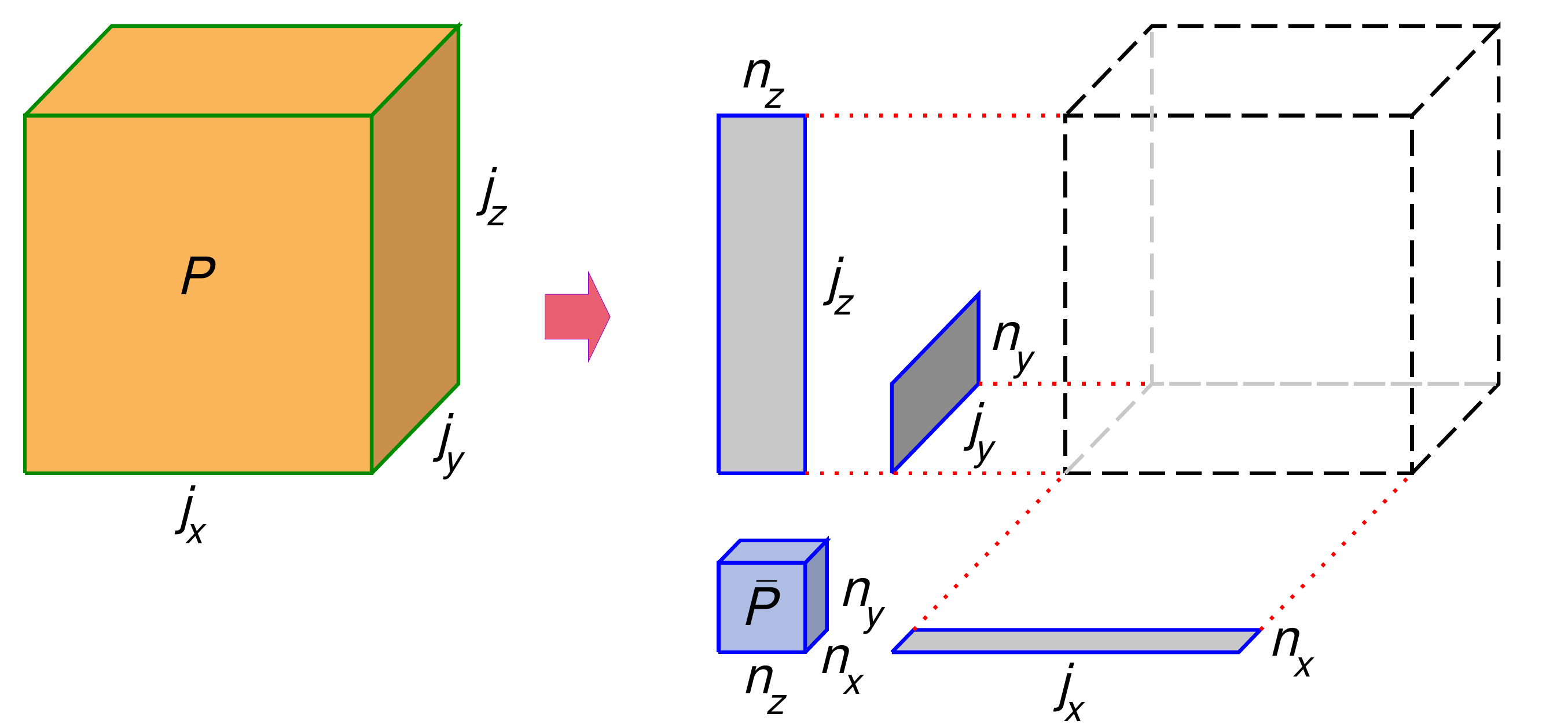}
    \caption{\label{fig:TuckerTensor}
    Rank-3 Tucker tensor decomposition.
	$P_{j_x j_y j_z}$ is compressed into
	$v^{[x]}_{n_x j_x}$, $v^{[y]}_{n_y j_y}$, $v^{[z]}_{n_z j_z}$
	and $\bar P_{n_x n_y n_z}$, see eq.~(\ref{eqn:TuckerTensor}).
    }%
   	\Description{
   	Decomposition of a tensor. The bulk 3D cube is decomposed into a
   	smaller cube times the Cartesian outer product of three sets of
   	vectors.
   	}%
\end{figure}
as depicted in fig.~\ref{fig:TuckerTensor}.
If it is possible to find such an approximation
with reduced ranges of the indices $(n_x,n_y,n_z)$ compared to the ranges of $(j_x,j_y,j_z)$
then $\bar P$ is a rank-3 tensor of much lower volume.
Furthermore, the total amount of memory needed for loading
the rectangular matrices $v^{[x|y|z]}$ in addition to loading $\bar P$
can be lower than the total volume of $P$,
depending on the original ranges and its compressibility.
In particular if we consider a set of projectors, as usual in PAW,
that can make shared use of the same 1D function sets $v^{[x|y|z]}$,
we expect an overall good compression ratio.

Transferring this scheme to the factorizable SHO basis, eq.~(\ref{eqn:SHO-factorizable-basis}),
we are seeking an approximation of the projector set $\tilde p_i(\vec r)$ of each atom as
\begin{equation}
 \tilde p_i(x,y,z) \approx \sum_{n_x n_y n_z} \bar p_{i\, n_x n_y n_z} \, \psi_{n_x}(x) \, \psi_{n_y}(y) \, \psi_{n_z}(z)
\end{equation}
with small ranges for $(n_x,n_y,n_z)$.

Dealing with compressed representations,
an additional decompression operation is necessary.
However, executing a task that is limited by the memory BW
usually implies idle arithmetic units in the processor,
so that the decompression phase is expected not add to the total execution time.

\subsection{SHO Basis}
Consider injecting the SHO basis projector $\hat {\mathcal B}_{\numax}$ as defined in eq.~(\ref{eqn:SHO-basis}),
but for a finite $\numax$ and given $\sigma$,
into the projection operation: $\braketop{ \tilde p_j }{ \hat {\mathcal B}_{\numax} }{ \Psi_k }$.
For simplicity of the notation we suppress the atom index $a$ here.
Then, we can evaluate the inner products
\begin{equation}
	\braket{ \tilde p_j }{ \nrn \ell m } \coloneqq F_{j \nrn \ell m}
\end{equation}
on radial grids during the initialization phase of the DFT calculation.
The operation left to be performed every time
is a projection onto the Cartesian factorizable SHO basis introduced in sec.~\ref{sec:SHO}:
\begin{align}
  C_{n_x n_y n_z k} &= \braket{ n_x n_y n_z }{ \Psi_k } \\
                    &= \iiint\limits_{\mathbb R^3} \mathrm d^3 \   \vec r
                       \, \psi_{n_x}(x) \, \psi_{n_y}(y) \, \psi_{n_z}(z) \, \Psi_k(x,y,z)
\end{align}

\noindent
In order to retrieve the original projection coefficients $c_{jk}$
we can transform the new coefficients $C_{n_x n_y n_z k}$
with $\braket{ \tilde p_j }{ \nrn \ell m } \equiv \hat F$
and $\braketR{\nrn \ell m}{n_x n_y n_z} \equiv \hat U$.
However, there is no need to retrieve these coefficients as
these are temporary quantities existing only for the duration of projection and expansion operations.
In fact, we can inject $\hat {\mathcal B}^\dagger_{\numax}$ also into the expansion operation.
Then, we collect all terms such that the original dyadic operator is fully transformed:
\begin{align}
 & \ket{ \tilde p_i } D_{ij} \bra{ \tilde p_j }
  \stackrel{ \text{\tiny{SHO}} }{ \longrightarrow }
  \  \  \hat {\mathcal B}^{\dagger}_{\numax} \, \ket{ \tilde p_i } \, D_{ij} \, \bra{ \tilde p_j } \, \hat {\mathcal B}_{\numax} \\
 &\phantom{:}= \ket{ n_x n_y n_z } \  U^{n_x n_y n_z}_{\nrn \ell m} \  F_{i \nrn \ell m}
               \  D_{ij} \ F_{j \nrn' \ell' m'} \   U^{\nrn' \ell' m'}_{n'_x n'_y n'_z}
               \bra{n'_x n'_y n'_z}
               \nonumber \\
 &\coloneqq    \ket{ n_x n_y n_z } \  {\mathcal D}^{n_x n_y n_z}_{n'_x n'_y n'_z} \bra{n'_x n'_y n'_z}
\end{align}
with contraction over adjacent indices (Einstein notation). \\
The atomic Hamiltonian term $\hat D$ is now replaced by
$\hat {\mathcal D}$ which can be computed as $\hat U^\dagger \hat F^\dagger \hat D \hat F \hat U$ at initialization of each SCF cycle.
$\hat {\mathcal D}$ represents a real-valued matrix of dimension $N\um{SHO}$.
See~eq.~(\ref{eqn:SHO_basis_size}) for the definition of $N\um{SHO}$.
%
\subsection{Smoothness}
Another advantage of the SHO basis for projectors is that it does not
require any filtering in reciprocal space
before it can be represented on a uniform Cartesian grid, c.f.~sec.~\ref{sec:Filtering}.
\\
In fact,
the highest kinetic energy in a 1D Hermite function
is given by its eigenenergy $E_{1D}=(2n$+$1)\,\sigma^{-2}$.
According to the sampling theorem, $k_{\max} \cdot h_{\max} = \pi$,
it is possible to sample the 1D Hermite functions
on a uniform real-space grid with maximum grid spacing
\begin{equation}
  h_{\max} = \frac{ \pi \, \sigma }{ \sqrt{2\numax + 1} }
  \label{eqn:max_grid_spacing}  \text{.}
\end{equation}
\\
Since Hermite functions are eigenfunctions of the Fourier transform
their representation in reciprocal space reads
\begin{equation}
  \mathcal F\left\{ H_n(x)\,\exp(-\frac{x^2}{2}) \right\}(k) \propto {(-1)}^n \, H_n(k) \, \exp(-\frac{k^2}{2})
  \text{.}
\end{equation}
The Gaussian suppresses high Fourier components in reciprocal space
and, thus, controls the smoothness of the function set.
We show in the appendix sec.~\ref{sec:Bessel-transform-eigenfunctions}
that the radial SHO eigenstates $R_{\nrn \ell}(r)$ are eigenfunctions of the Fourier-Bessel transform.
Hence, the radial SHO eigenstates also feature a Gaussian decay in Fourier space.
This shows that SHO states are the simultaneously best localized functions in both spaces.

\subsection{On-the-fly Basis}
Using the recursion relation for Hermite polynomials
\begin{equation}
  H_0(x) = 1, \  \  \    H_1(x) = x, \  \  \   H_{n+1}(x) = x \, H_n(x) - \frac{n}{2} \, H_{n-1}(x)
\end{equation}
it is even possible to cheaply evaluate the grid-sampled 1D Hermite functions
during the decompression phase.
This reduces the total transferred memory volume
to a few Bytes.
Hence, it effectively eliminates the memory BW requirement
for the data list of the sparse matrix
that results from the projector functions.
The missing factors ensuring normalization of the Hermite functions
can be absorbed into the transformed atomic matrices $\hat {\mathcal D}$.

We might experience that in some situations the SHO basis
needs to be considerably larger than the original set of numerically given projectors
which increases the BW requirements for the intermediate coefficients $C_{n_x n_y n_z k}$
for each atom
that have to be stored after projection and loaded before expansion.
Therefore, we investigate on the necessary minimum size of a SHO basis
that still yields an acceptable representation quality of the PAW projector functions.

\section{Quality of Representation}\label{sec:QLT}
We investigate how well commonly used projector functions $P_{n\ell}(r)$ can
be represented in the radial SHO basis $R_{\nrn \ell}(r/\sigma)$
as a function of the spread $\sigma$.
We define the representation quality $Q$ as
\begin{equation}
	Q_\nu(\sigma) = \sum_{\nrn = 0}^{ \lfloor \frac 12(\nu - \ell) \rfloor }
	\left| \braket{ P_{n\ell}(r) }{ R_{\nrn \ell}(r/\sigma) } \right|^2
	\text{.}
\end{equation}
Applying this metric to all PAW data sets (\ttt{*.PBE})
publicly available in the package \ttt{gpaw-setups-0.9.9672}
from GPAW~\cite{GPAW-website}
shows that we can represent most unfiltered projector functions
with at least \unit[90]{\%} quality. 
Even higher values are seen
if the projector functions are mask-filtered~\cite{doi:10.1063/1.2193514} before.

\begin{figure}
	\centering
	\includegraphics[width=.92\linewidth]{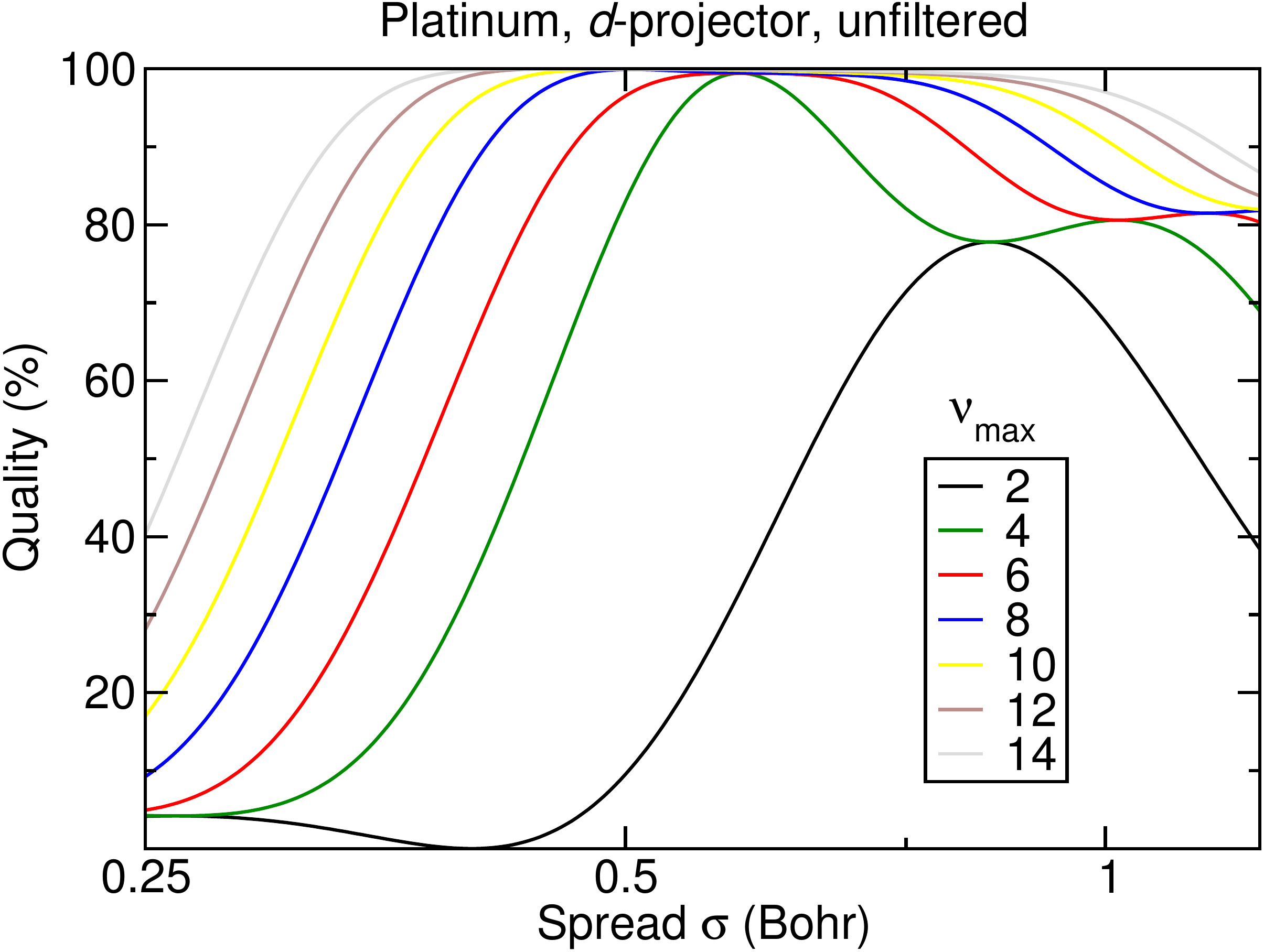} 
    \caption{\label{fig:quality_convergence_Pt_d_raw}
    Convergence of the representation quality
    for the $d$-projector of platinum.
    Up to seven radial SHO basis functions were used ($\numax = 14$).
    With respect to the size of the SHO basis,
    the strongest increase is between $\numax = 2$ and $4$.
	}%
   	\Description{
   	The maxima become broader the more basis functions are added.
   	}%
\end{figure}
A typical convergence of $Q$ w.r.t.~$\numax$ can be observed
at the example of the platinum (Pt) $d$-projector,
see fig.~\ref{fig:quality_convergence_Pt_d_raw}.
For $\numax = 2$, the radial SHO basis in the $d$-channel consists of a single function
that features no radial nodes.
The Pt-$d$-projector function from the GPAW data base
can only be represented up to $Q = \unit[77]{\%}$ at $\sigma = \unit[0.84]{Bohr}$ (solid black line).
Already for two basis function ($\numax = 4$),
we can reach up to $Q = \unit[99.4]{\%}$ at $\sigma = \unit[0.59]{Bohr}$ (solid green line).
Adding more basis functions does not increase $Q$ much
but makes the maxima wider, i.e.~a larger basis is capable of accurately representing
the numerically given projectors with some flexibility on $\sigma$.
We can see that the best increase in quality is reached with $\numax = 4$.
For higher $\numax$, the gain in $Q$ is small compared to the increase in costs
since the SHO basis size grows $\propto \numax^3$, c.f.~eq.~(\ref{eqn:SHO_basis_size}).

The simplicity of the SHO basis is an advantage
and a drawback at the same time.
Only a single $\sigma$ can be selected for all projector functions of one atom.
\begin{figure}
	\centering
	\includegraphics[width=.92\linewidth]{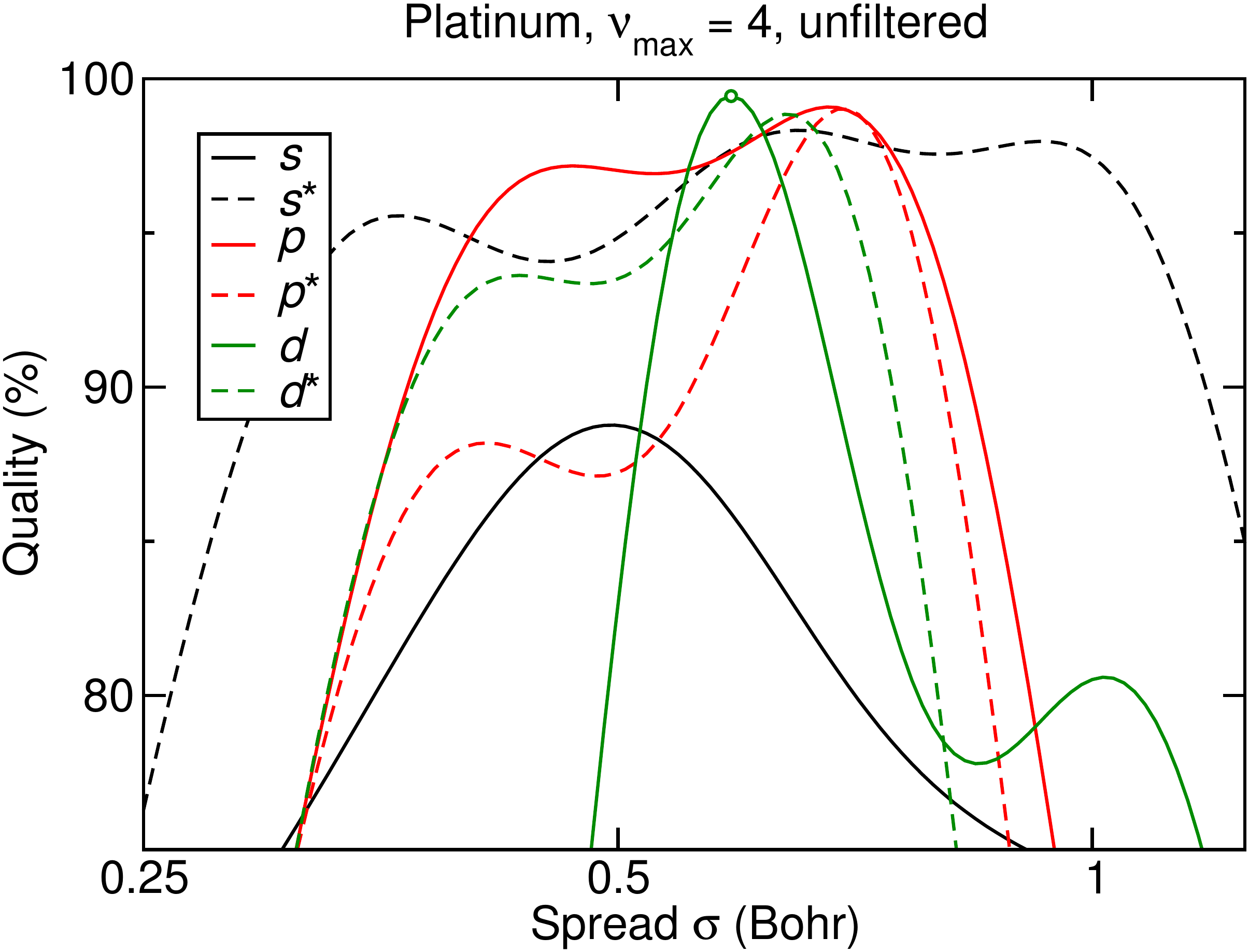} 
   	\caption{\label{fig:quality_Pt_raw}
  	Best representation quality as a function of $\sigma$ for
   	all projectors of platinum with $\numax = 4$.
	Colors are chosen according to $\ell$-character,
	solid and dashed lines show the first and second~(*) projector, respectively.
	We select $\sigma = \unit[0.59]{Bohr}$ (green dot).
	The solid green line again shows the quality of the Pt-$d$-projector as above. 
	}%
   	\Description{
   	The peaks of the quality curves are at different $\sigma$.
   	We manually select the narrowest curve.
   	}%
\end{figure}
As visible in fig.~\ref{fig:quality_Pt_raw} the quality functions for
the $s$, $s^*$, $p$, $p^*$, $d$ and $d^*$-projectors
have their maxima at slightly different values of $\sigma$.
Therefore, we selected the peak of the sharpest curve
which for the case of platinum is the $d$-projector.
As discussed above, the best quality for $\numax = 4$
is found at $\sigma = \unit[0.59]{Bohr}$ which is marked with a dot in fig.~\ref{fig:quality_Pt_raw}.

With respect to costs, we would like to keep the SHO basis as small as possible.
The minimal number of radial basis functions can be derived
from the number of projectors in each $\ell$-channel.
The minimal $\numax$ for all PAW data sets analysed can be found in the appendix sec.~\ref{sec:minimum_nu_max}.
This is, as a rule of thumb, $\numax = 4$ for transition metals,
$\numax = 3$ for non-metals and $\numax = 2$ for light elements, $Z < 5$.
Further, we suggest $\numax = 5$ for the treatment of rare earth elements
in order to host two projectors in the $f$-channel.
For some elements with the minimal $\numax$
it is, however, difficult to find a single value $\sigma$
that produces good representation qualities for all projectors.
As it would be desirable in terms of costs to be able to use the minimal $\numax$
we suggest a modified PAW dataset generation procedure described in the following section.

\section{Modified PAW data generation}\label{sec:MOD}
\begin{figure}
	\centering
	\includegraphics[width=\linewidth]{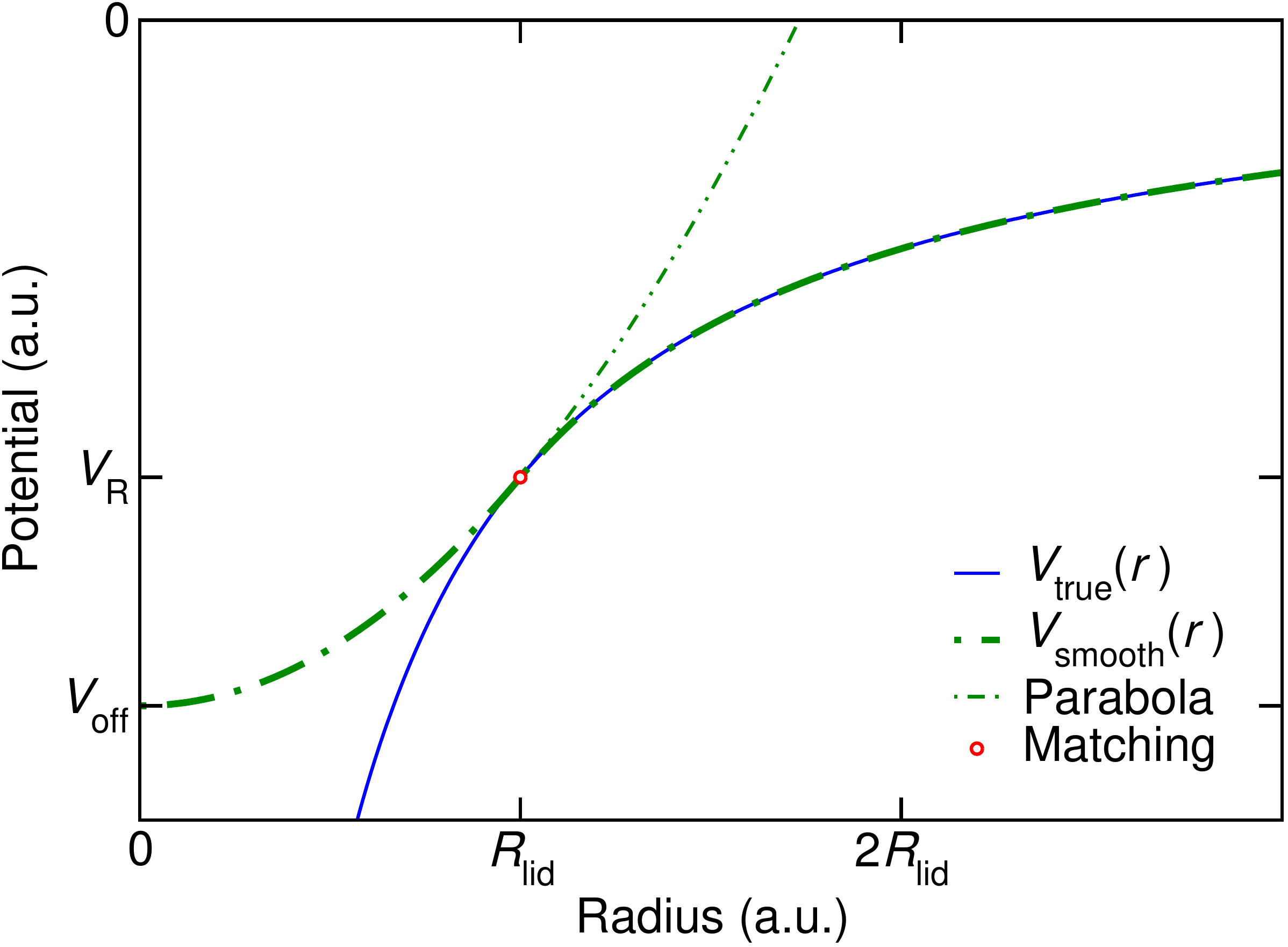} 
   	\caption{\label{fig:potential-lid-technique}
  	Pseudization of the local potential according to the \emph{potential lid technique}.
  	A parabola of given curvature closes the true deep core potential like a lid closes a pot.
	}%
   	\Description{
   	A convex parabola is touching a concave potential function with equal tangent.
   	}%
\end{figure}

For the generation of PAW data sets several
different recipes can be found in the literature,
see e.g.~\cite{JOLLET20141246} or~\cite{PhysRevB.59.1758} and references therein.
This starts at the pseudization of the true local potential
$V\um{eff}(r) \rightarrow \tilde V\um{eff}(r)$
i.e.~there are different shapes for a smooth continuation of the potential inside the augmentation sphere.
For the generation of true partial waves $\phi_{n \ell}(r)$,
a set of energy parameters $\varepsilon_{n \ell}$ needs to be chosen.
Assuming a spherically symmetric true potential $V\um{eff}(r)$,
a true partial wave $\phi_{n\ell}(r)$ is found by outwards integration
of the radial ordinary differential equation (ODE) for a given energy $\varepsilon_{n \ell}$.
This can be either the scalar relativistic equation or the Schr\"{o}dinger equation.
Furthermore, the generation of smooth partial waves $\tilde \phi_{n \ell}(r)$ and
projector functions $\tilde p_{n \ell}(r)$ depends on the recipe one follows.
The simplest prescription is to use $r^\ell$ times a low-order polynomial in $r^2$
for the construction of the smooth partial waves, $\tilde \phi_{n \ell}(r)$.
This is matched with value and derivative
to the true partial wave $\phi_{n\ell}(r)$ at $r$=$R\um{aug}$~\cite{Rostgaard:2009:PAW-Note}.
Once obtained $\tilde \phi_{n \ell}(r)$, a first guess for the shape
of the radial part of the projector function comes from
\begin{equation}
  \ket{ \tilde p_{n \ell} } =
  \left( \hat T + \tilde V\um{eff} - \varepsilon_{n \ell} \right)
  \ket{ \tilde \phi_{n \ell}}
  \label{eqn:usual-projector-generation}
\end{equation}
as suggested by Bl\"ochl in his original work on PAW~\cite{PhysRevB.50.17953}.
In order to keep the SHO basis small (minimum $\numax$, if possible)
we suggest to invert this scheme.
We regard eq.~(\ref{eqn:usual-projector-generation}) as inhomogeneous
ODE and solve it for $\tilde \phi_{n \ell}(r)$
using the radial SHO eigenstates $R_{\nrn \ell}(r/\sigma)$ with $\nrn$=$n$ as projector functions:
\begin{equation}
  \ket{ \tilde \phi_{n \ell}} =
  {\left( \hat T + \tilde V\um{eff} - \varepsilon_{n \ell} \right)}^{-1}
  \ket{ R_{n \ell} }
  \label{eqn:suggested-projector-generation}
\end{equation}
This removes a lot of free parameters from the PAW generation scheme
as we are only left with $\numax$ and $\sigma$.
The cut-off radius $R\um{aug}$ is only required
for the construction of $\tilde V\um{eff}(r)$.
We can even try to eliminate this dependency:
GPAW constructs $\tilde V\um{eff}(r)$ as parabola for $r < R\um{aug}$
which matches the true potential at $R\um{aug}$ in value and first derivative.
If we constrain the curvature of this parabola to a fixed function of $\sigma$, e.g.~${(2\sigma)}^{-4}$,
the parabola and the true potential touch at a radius $R\um{lid}$
which comes out of the procedure instead of being an input.
We call this the \emph{potential lid technique}
since the parabola's offset is lowered until it touches the true potential,
much like a lid that closes a pot.
See fig.~\ref{fig:potential-lid-technique} for a schematic picture.
The factor of two rescaling $\sigma$ has been chosen to produce good results for iron.
However, we suspect it to work well for a wider range of elements
which would allow to eliminate it from the list of tunable parameters.

As we want the pseudo Hamiltonian to produce the same energy ordering in the valence range
as the true Hamiltonian, we demand that
that $\varepsilon_{0 \ell} < \varepsilon_{1 \ell} < \cdots < \varepsilon_{n_{\max, \ell} \ell}$
where $n_{\max, \ell} = \left\lfloor \frac 1 2 (\numax - \ell) \right\rfloor$.

\subsection{Results}\label{sec:results}
As an example, we will discuss the results obtained for iron ($Z = 26$).
Here, we find a lid radius $R\um{lid}$ of \unit[2.08]{Bohr}
which is very close to the radius used in the GPAW data set.
A spread parameter of $\sigma = \unit[0.65]{Bohr}$ for the analytical SHO projectors
was selected to yield the simultaneously best representation quality
of all projectors with $\numax = 4$.
%
\begin{figure}
	\centering
	\includegraphics[width=\linewidth]{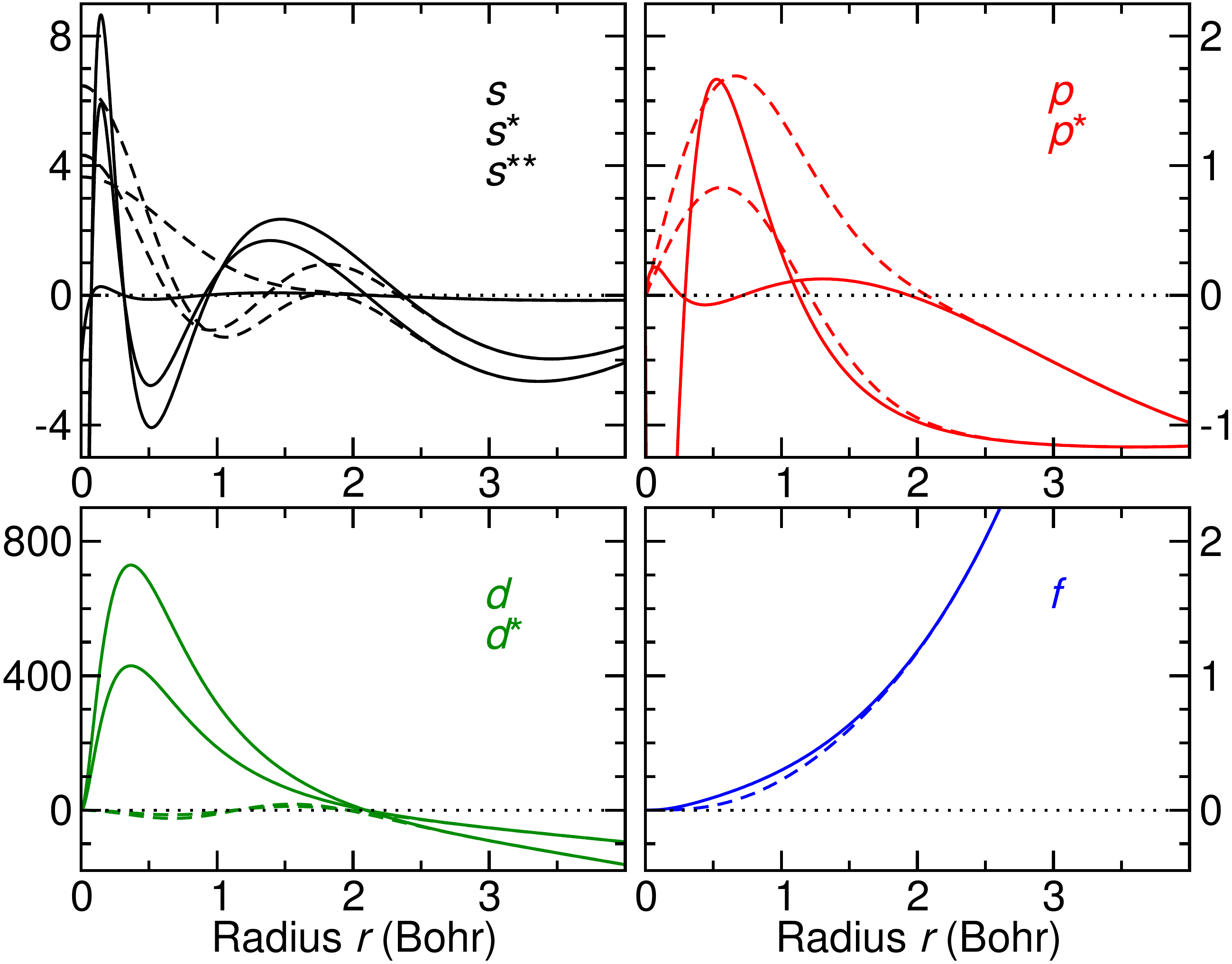} 
    \caption{\label{fig:partial-waves-for-Fe}
    Iron true partial waves (solid lines) and smooth partial waves (dashed lines)
    generated by the inverted PAW generation scheme.
	}%
   	\Description{
   	In the $s$-channel, true partial waves feature rapid oscillations close to the core.
   	In the $d$-channel, the true partial waves show the Fe-3$d$-states localized mostly inside $r \leq 2\,$Bohr.
   	}%
\end{figure}
The selected reference energies are $-0.4$ (Fe-4$s$), $-0.11$ (Fe-4$p$)
and \unit[-0.57]{Rydberg} (Fe-3$d$). The latter is also used for $\ell > 2$.
Additional true partial waves are generated at $\frac{n}{\ell + 1}\,$Rydberg
above their reference energies.
The resulting true and smooth partial waves are shown in fig.~\ref{fig:partial-waves-for-Fe}.

\begin{figure}
	\centering
	\includegraphics[width=\linewidth]{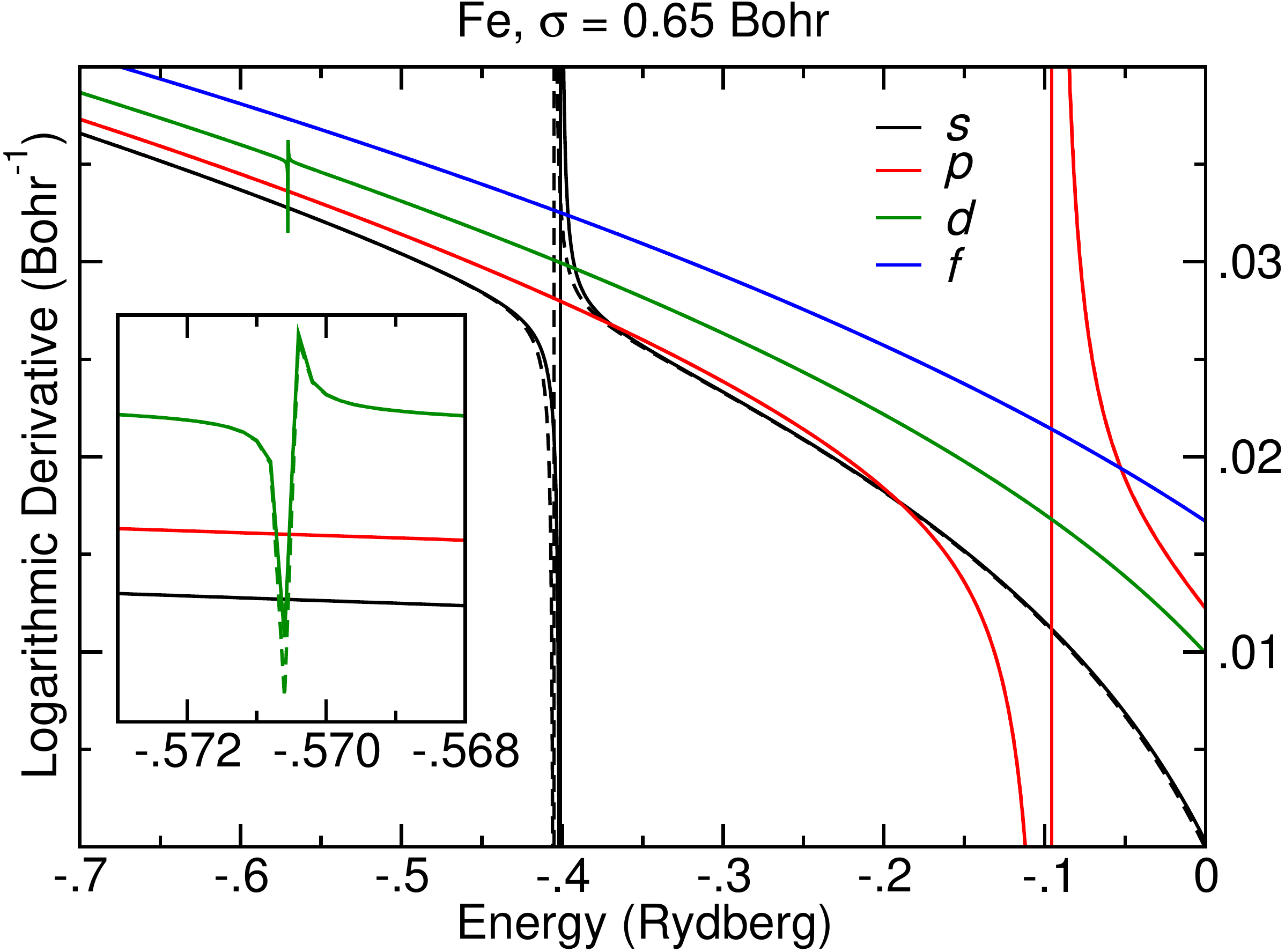} 
    \caption{\label{fig:logder-for-Fe}
    Logarithmic derivative (at $r$=$\unit[7.1]{Bohr}$) generated by the inverted PAW generation scheme for iron.
	The strongest deviation inside the valence energy range
	is a \unit[4]{mRydberg} shift of the resonance in the $s$-channel.
	All other dashed lines (pseudo logarithmic derivatives) are covered by the solid lines (true logarithmic derivatives).
	}%
   	\Description{
   	All logarithmic derivative are falling curves with poles at
   	energies where resonances appear. Their shape is similar to that of a tangent.
   	}%
\end{figure}

We verify the scattering properties of the new PAW dataset for iron
generated according to eq.~(\ref{eqn:suggested-projector-generation})
by comparing its logarithmic derivative $L_\ell(E)$ to that of the true potential,
see fig.~\ref{fig:logder-for-Fe}.
Except for a \unit[4]{mRydberg} shift in the position of the $s$-resonance,
the (solid) lines of the true $L_\ell(E)$ lie on top of the pseudo $L_\ell(E)$ (dashed lines)
which is a reasonably good agreement.
The Fe-3$d$-states are strongly localized which makes them appear
as a narrow resonance at \unit[-0.57]{Rydberg}, see inset in fig.~\ref{fig:logder-for-Fe}.
Since the spread $\sigma$ and the cutoff $\numax$
are tunable input parameters to the generation procedure
we have sufficient freedom to produce a transferable PAW dataset.

\subsection{Rescaling of partial waves}
Bl\"{o}chl's PAW generation scheme~\cite{PhysRevB.50.17953} 
allows to rescale partial waves and projectors according to
\begin{equation}
               \left(    \phi_{n\ell},     \tilde \phi_{n\ell},       \tilde p_{n\ell} \right)
  \rightarrow  \left( s\,\phi_{n\ell}, s\, \tilde \phi_{n\ell}, s^{-1}\tilde p_{n\ell} \right)
  ,\  \  \   s \in \mathbb R \setminus \{0\}
  \label{eqn:PAW_partial_waves_rescaling}
\end{equation}
independently for each $n$ and $\ell$.
This is because after a successful PAW generation procedure, the dual orthogonality
\begin{equation}
	\braket{ \tilde p_{n\ell} }{ \tilde \phi_{n'\ell} } = \delta_{nn'}
	\label{eqn:dual_orthogonality_of_smooth_partial_waves_and_projectors}
\end{equation}
must hold.
Typically, this orthogonality is established by rescaling
the lowest projector and the lowest partial waves
and orthogonalizing the higher ones to the lowest.
This procedure is equivalent to an $LU$-decomposition.

With the SHO projectors, rescaling is not allowed
if we aim to exploit the 3D factorization,
c.f.~eq.~(\ref{eqn:SHO-transform-Cartesian-to-radial}).
It is rather important that the $R_{n\ell}(r/\sigma)$ enter eq.~(\ref{eqn:suggested-projector-generation})
with proper normalization.
Therefore,
eq.~(\ref{eqn:dual_orthogonality_of_smooth_partial_waves_and_projectors})
is enforced by applying the inverse of $\braket{ \tilde p_{n\ell} }{ \tilde \phi_{n'\ell} }$
only to the preliminary pairs of true and smooth partial waves.

\section{Performance Measurements}\label{sec:PERF}
\begin{figure}
	\centering
	\includegraphics[trim=3cm 3cm 3cm 3cm,clip,width=.5\linewidth]{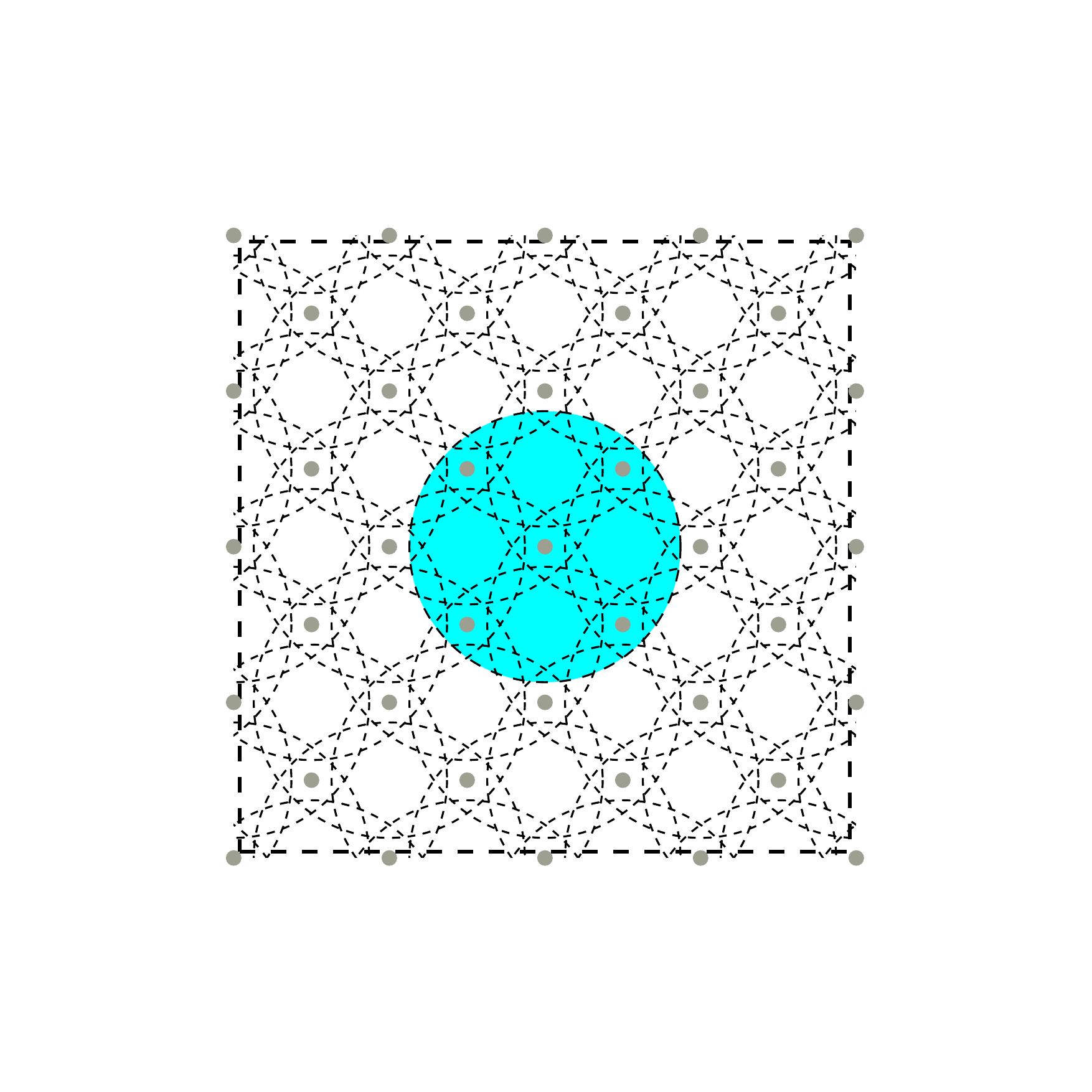} 
    \caption{\label{fig:gold_projection_radii}
    2D sketch of the overlapping projection spheres inside a cubic domain of edge length \unit[16]{\AA}.
    This setup is used for the GPU performance benchmarks in \ttt{double} precision.
	In average, the wave function values of each grid point
	are involved into eleven projection operations.
	The radius of each projection sphere is \unit[3.55]{\AA}
	while the nearest neighbor distance of atomic nuclei is \unit[2.9]{\AA}.
	}%
   	\Description{
   	Test system for performance benchmark
   	}%
\end{figure}

%
For an impression of the expectable savings
we assess the performance of projection (\texttt{prj}) and expansion (\texttt{add})
operations as defined in sec.~\ref{sec:HMT} by (1) and (3), respectively,
for a first implementation on NVIDIA GPUs.
In order to provide a meaningful comparison between
the runtime of \texttt{SHO} projectors which are generated on-the-fly
and the reference (\texttt{USU}) using precomputed projector values,
we work in the same framework and exchange only the kernel.
The test system is a cubic domain of edge length \unit[16]{\AA}
with a grid spacing of \unit[0.25]{\AA}, i.e.~$64 \times 64 \times 64$ grid points.
Assuming a face-centered cubic crystal of atoms with a
lattice constant of \unit[4.08]{\AA} (e.g.~silver or gold) 
lets us find $241$ atom positions inside the domain. 
With a projection radius of \unit[3.55]{\AA} exactly $665$ atoms contribute with a non-zero overlap of
their projection sphere
and the cubic domain,
see fig.~\ref{fig:gold_projection_radii}.
Mind that there is also a non-zero overlap between projection spheres of
neighboring atoms. Unlike the augmentation spheres,
the projection spheres may overlap.


The Hamiltonian is applied to $1024$ wave functions at a time.
This index is used for vectorization over warps of $32$ GPU-threads.
The performance results are converged w.r.t.~this number.

For stable performance results the kernel execution is repeated $10$ times.
After a warm-up run which is not considered,
the median of $15$ runs is reported in tab.~\ref{tab:perf-results}.

The memory consumption for the precomputed projectors is
\unit[76]{MByte} per projection coefficient.
This results in an additional GPU memory request of up to \unit[4.2]{GByte}.


We executed on an {IBM PowerNV 8335-GTB}
with {NVIDIA Tesla P100-SXM2} (Pascal) GPUs connected by \ttt{NVlink}.
Runtime and compilers are \ttt{CUDA 9.2.148}, \ttt{GCC 7.2.0}.
Furthermore, we benchmarked the latest {NVIDIA Tesla V100} (Volta) GPUs
mounted on an Intel system (Xeon Gold 6148) connected via PCIe.
Here, \ttt{CUDA 9.2.148} and \ttt{GCC 7.3.0} toolchains were used.

Estimating the upper limit of floating point operations executed
we can verify that all kernels are performance-limited by the device memory BW.
Hence, an improvement in the timings directly relates to savings in the BW.

From tab.~\ref{tab:perf-results} we can see an up to $6.6$ times faster execution
depending on the kernel, the basis size and the device used.
The speedup $S$ compares the timings of the
\ttt{USU}al projection operations using precomputed projector functions
to timings of the \ttt{SHO} projection operations.
While \ttt{USU} kernels can take any number of projectors
a SHO basis of size $18$ cannot be constructed, therefore,
the corresponding entries for \ttt{SHO} are empty.
However, we included the timing result for \ttt{USU18}
as it allows to consider a typical calculation setup for transition metal elements.
In order to cover $s$, $s^*$, $p$, $p^*$, $d$ and $d^*$ projectors,
the SHO basis must be constructed with $\numax = 4$, i.e.~$35$ projection
coefficients.
Hence, a fair comparison of the two methods is to take \ttt{USU18} vs.~\ttt{SHO35}
which translates into an algorithmic speedup
given in~tab.~\ref{tab:perf-fair-comparison}.
Executing both kernels after each other, \ttt{prj+add},
in \ttt{double} precision
results in an algorithmic speedup of $2.6$ on P100 GPUs
and $2.0$ on the more recent Volta architecture, V100.


%

\begin{table}
 \caption{
		Performance improvement on NVIDIA GPUs. All times are reported in seconds.
		The speedup $S$ is the unitless ratio of timings \ttt{USU}/\ttt{SHO} for the same \# of projectors.
 }
 \label{tab:perf-results}
 \begin{tabular}{r rrr rrr}
   \toprule
		   & \multicolumn{6}{c}{NVIDIA P100 Pascal} \\
      \#   & \ttt{USUprj} & \ttt{SHOprj} &     $S$ & \ttt{USUadd} & \ttt{SHOadd} &     $S$ \\
   \midrule 
	   1   &        0.480 &        0.453 &     1.1 &        0.274 &        0.187 &     1.5 \\
	   4   &        0.834 &        0.471 &     1.8 &        0.803 &        0.363 &     2.2 \\
	  10   &        1.259 &        0.430 &     2.9 &        1.820 &        0.520 &     3.5 \\
	  18   &        2.080 &              &         &        3.173 &              &         \\
	  20   &        2.366 &        0.482 &     4.9 &        3.511 &        0.733 &     4.8 \\
	  35   &        3.904 &        0.967 &     4.0 &        6.052 &        1.027 &     5.9 \\
	  56   &        6.058 &        1.214 &     5.0 &        9.625 &        1.455 &     6.6 \\
   \midrule 
		   & \multicolumn{6}{c}{NVIDIA V100 Volta} \\
      \#   & \ttt{USUprj} & \ttt{SHOprj} &     $S$ & \ttt{USUadd} & \ttt{SHOadd} &     $S$ \\
   \midrule 
	   1   &        0.291 &        0.271 &     1.1 &        0.140 &        0.118 &     1.2 \\
	   4   &        0.354 &        0.310 &     1.1 &        0.378 &        0.189 &     2.0 \\
	  10   &        0.555 &        0.284 &     2.0 &        0.823 &        0.277 &     3.0 \\
	  18   &        0.830 &              &         &        1.411 &              &         \\
	  20   &        1.013 &        0.300 &     3.4 &        1.557 &        0.398 &     3.9 \\
	  35   &        2.087 &        0.575 &     3.6 &        2.660 &        0.564 &     4.7 \\
	  56   &        1.806 &        0.793 &     2.3 &        4.208 &        0.792 &     5.3 \\
   \bottomrule
 \end{tabular}

\end{table}

\begin{table}
 \caption{
		Algorithmic speedup comparing \ttt{USU18} vs.~\ttt{SHO35}.
 }
 \label{tab:perf-fair-comparison}
 \begin{tabular}{r rrr rrr}
   \toprule
           & \multicolumn{3}{c}{\ttt{float}}    & \multicolumn{3}{c}{\ttt{double}}   \\
    GPU    & \ttt{prj} & \ttt{add} & \ttt{both} & \ttt{prj} & \ttt{add} & \ttt{both} \\
   \midrule 
	P100   &       2.3 &       2.9 &        2.7 &       2.2 &       3.1 &        2.6 \\
	V100   &       1.4 &       1.7 &        1.6 &       1.4 &       2.5 &        2.0 \\
   \bottomrule
 \end{tabular}




%


%


\end{table}

\section{Discussion}\label{sec:discuss}
%
The suggested application of a SHO basis for projection operations
or, even going further, analytical SHO projector functions
brings many advantages but also comes with some drawbacks.
In this section, we discuss pro and contra.

\subsection{Advantages}
The SHO basis
\begin{itemize}
 \item factorizes in the Cartesian coordinates $(x,y,z)$.
 \item can be sampled on uniform grids without filtering.
 \item has only two parameters, $\sigma$ and $\numax$.
 \item is given analytically.
\end{itemize}
In particular the last point, the analytical shapes,
allows to save memory bandwidth and memory capacity
for implementations of grid-based projection operations,
as shown in sec.~\ref{sec:PERF}.

\subsection{Drawbacks}
The SHO basis does not offer the flexibility
to represent any projector function with good quality.
However, we have shown that the PAW generation scheme
can be adjusted to using radial SHO basis as projectors.

The SHO basis leads to more projection coefficients than usual PAW.\@ 
For example, two projectors for the $s$, $p$ and $d$-channel makes $18$ coefficients.
In order to fit a $d^*$-projector into a SHO basis,
we need a minimum $\numax$ of $4$, i.e.~$35$ coefficients.
However, we have to take it as an upside.
With $\numax = 4$ the SHO basis contains
an additional $s^{**}$, seven $f$ and nine $g$-projectors.
Adding projectors and partial waves
of higher $\ell$
may potentially improve the transferability of the PAW data sets
for chemical environments with low symmetry
since gradients in the potential scatter into the next higher $\ell$-channel.
Compared to commonly used projector sets, SHO projectors treat also higher $\ell$-channels in full-potential accuracy. 
However, the true  transferability remains to be shown in 3D calculations.

\section{Related work}\label{sec:related}
%
\noindent
Already 1986, Obara and Saika presented the ansatz of 3D products of
primitive Cartesian Gaussian (pCG) functions
$x^{\nu}\exp(-\frac 12 x^2)$ 
in order to evaluate two- and three-center integrals~\cite{doi:10.1063/1.450106}.
In fact, there is a strong relation between pCGs and the 1D harmonic oscillator states.
The pCG functions form a non-orthogonal basis set.
Applying Gram-Schmidt orthogonalization to pCGs leads to 1D Hermite functions.
\noindent
Many works in quantum chemistry,
as e.g.~Schlegel and Frisch~\cite{doi:10.1002/qua.560540202},
have mentioned the increased basis size $\frac 12(\lmax + 2)(\lmax + 1)$
of the pCGs compared to $(2\lmax + 1)$
but this has usually been taken as a weakness rather than a strength.

The idea of using the SHO basis has probably not received attention as the related basis set
\begin{equation}
	 \chi_{\nrn \ell m}(\vec r) = r^{\ell + 2\nrn} \, \exp(-\frac {r^2}{2\sigma^2}) \   Y_{\ell m}(\hat r)
\end{equation}
has been found not to converge as rapidly as a set of only Gaussian-type orbitals (GTOs)
contracted over a set of different spread parameters $\sigma$~\cite{doi:10.1002/jcc.540070403}
when used as a basis set for quantum chemistry calculations~\cite{doi:10.1063/1.463096}.
\\
Note that the subset of radial SHO states with $\nrn = 0$ are GTOs.

\section{Conclusions}\label{sec:summary}
The eigenstates of the Spherical Harmonic Oscillator (SHO)
form a natural link between 3D Cartesian grids
and radial representations with sharp quantum numbers for the angular momentum, $(\ell,m)$,
without the use of global Fourier transforms.
Projection and expansion operations
are the heart of the Projector Augmented Wave (PAW) method~\cite{PhysRevB.50.17953} for
electronic structure calculations.
Using the SHO basis for these operations
allows to exploit its special properties.
Similar to compressed tensor representations,
the 3D SHO states can be factorized into eigenstates of the 1D harmonic oscillator
for each of the three Cartesian coordinates $(x,y,z)$.
This offer a great potential for
reducing memory bandwidth requirements and memory capacity constraints.
In particular since it is cheap to recompute
the analytically given 1D basis functions
compared to loading them from memory.
A first implementation on GPUs shows that we can expect
an at least two times faster application of the non-local parts
of the PAW Hamiltonian for transition metals.
Commonly used PAW projector functions can be represented in a SHO basis,
however, the cost-saving minimal basis size is not suitable for representing arbitrary functions.
Therefore, we suggest a modified PAW generation scheme
in which the projector functions are fixed to the analytically given radial SHO basis functions
eliminating the freedom of choosing the shape and scale of the smooth partial waves.

\section*{Acknowledgements}
The authors thank Thorsten Hater for proofreading
and Miriam Hinzen for assistance with the Fourier-Bessel transform.


%

%
\bibliographystyle{ACM-Reference-Format}
\bibliography{literature}

%
\appendix

\section{Minimum $\numax$ for all elements}\label{sec:minimum_nu_max}
We suggest the minimal $\numax$ for the GPAW setups
in the structure of a periodic table of elements
from $_1$H to $_{86}$Rn:
\begin{Verbatim}[samepage=true]
     2                                         2
     2 2                             3 3 3 3 3 3
     3 3                             3 3 3 3 3 3
     3 3 4         4 4 4 4 4 4 4 4 4 3 3 3 3 3 3
     3 3 4         4 4 4 - 4 4 4 4 4 4 4 4 3 3 3
     3 4 4 5 ... 5 4 4 4 4 4 4 4 4 4 4 4 4 - - 3
\end{Verbatim}
Rare earth elements (RE) are not included in the publicly available collection \ttt{gpaw-setups-0.9.9672}~\cite{GPAW-website}.
We suggest $\numax = 5$ for REs in order to have two projectors in the $f$-channel.
For $\numax = 2$, $3$, $4$ or  $5$ the size of a
3D SHO basis is  $10,  20,  35$ or $56$, respectively.

\section{Eigenfunctions of the Fourier-Bessel transform}\label{sec:Bessel-transform-eigenfunctions}
The radial SHO eigenfunctions
shown in fig.~\ref{fig:radial_SHO_states_by_ell}
are eigenfunctions of the Fourier-Bessel transform ($\mathcal{FB}$)
which reads
\begin{equation}
  \mathcal{FB}\{R_{n\ell}(r)\}(q) = \sqrt{\frac 2 \pi} \int_{0}^{\infty} r^2 \mathrm d r \, j_\ell(qr) \  \   R_{n\ell}(r)
\end{equation}
where $j_\ell(x)$ are the spherical Bessel functions of the first kind featuring a regular $x^\ell$ behaviour at the origin.
The proof is given by Brackx \emph{et al.}~\cite{Brackx2009TheFT} (dimension $m = 3$, degree $k = 0$)
exploiting the identity given by Magnus \emph{et al.}~\cite{Magnus1966}, page~244: 
\begin{equation*}
 \int_0^\infty \mathrm d r \exp(-\frac{r^2}2) r^{\alpha + 1} L_{n}^{(\alpha)}(r^2) J_{\alpha}(r\sqrt{x})
 = {(-1)}^n x^{\alpha/2} \exp(-\frac x 2) L_{n}^{(\alpha)} (x) \nonumber \text{.}
\end{equation*}
Replacing $\alpha = \ell + \frac 12$ and $x = q^2$ this becomes
\begin{align*}
 \int_0^\infty \mathrm d r \, J_{\ell + \frac 12}(qr)
            \, r^{\ell + \frac 32} \, &\exp(-\frac{r^2} 2) \, L_{n}^{(\ell + \frac 12)}(r^2) \\
 = {(-1)}^n \, q^{\ell + \frac 12} \, &\exp(-\frac{q^2} 2) \, L_{n}^{(\ell + \frac 12)}(q^2) \nonumber
 \text{.}
\end{align*}
With the common definition of spherical Bessel functions
\begin{align*}
 j_{\ell}(z) = \sqrt{\frac{\pi}{2z}} \, & J_{\ell + \frac 12}(z) \nonumber
 \qquad \text{we arrive at} \\
 \sqrt{\frac 2 \pi} \int_0^\infty r^2  \mathrm d r \,  j_{\ell}(qr) \   \   r^{\ell} &\exp(-\frac{r^2} 2) \, L_{n}^{(\ell + \frac 12)}(r^2)  \\
 = {(-1)}^n \, q^{\ell} &\exp(-\frac{q^2} 2) \, L_{n}^{(\ell + \frac 12)} (q^2) \nonumber \text{.}
\end{align*}
This shows that the $\mathcal{FB}$-transform produces the same function with exchanged argument ($r \leftrightarrow q$) and eigenvalue ${(-1)}^n$.

\end{document}